\documentclass[ams, prb, twocolumn, amssymb, floatfix, longbibliography]{revtex4-2}
\usepackage{amsmath}
\usepackage{tabularx}
\usepackage{bm}
\usepackage{euscript}
\usepackage{graphicx}
\usepackage[usenames,dvipsnames]{xcolor}
\usepackage{amsfonts}
\usepackage{exscale}
\usepackage{amsbsy}
\usepackage{subfigure}
\usepackage{textcomp}
\usepackage{comment}
\usepackage{slashed}
\usepackage{hyperref}
\usepackage{braket}
\usepackage{multirow}
\usepackage{array}

\hypersetup{
	colorlinks=true,
	linkcolor=black,
	urlcolor=cyan,
	citecolor=ForestGreen
}

\newcommand{\beq}{\begin{equation}}
\newcommand{\eeq}{\end{equation}}
\newcommand{\be}{\begin{eqnarray}}
\newcommand{\ee}{\end{eqnarray}}

\begin{document}
	


\title{Scaling of entanglement entropy at quantum critical points in random spin chains}
\author{Prashant Kumar$^{1,2}$ and R. N. Bhatt$^{3}$}
\affiliation{
$^1$Department of Physics, Princeton University, Princeton NJ 08544, USA\\
$^2$Kadanoff Center for Theoretical Physics, University of Chicago, Chicago, Illinois 60637, USA\\
$^3$Department of Electrical and Computer Engineering, Princeton University, Princeton NJ 08544, USA
}
\date{\today}

\begin{abstract}
	We study the scaling properties of the entanglement entropy (EE) near quantum critical points in interacting random antiferromagnetic (AF) spin chains. Using density-matrix renormalization group, we compute the half-chain EE near the topological phase transition between Haldane and Random Singlet phases in a disordered spin-1 chain. It is found to diverge logarithmically in system size with an effective central charge $c_{\rm eff} = 1.17(4)$ at the quantum critical point (QCP). Moreover, a scaling analysis of EE yields the correlation length exponent $\nu=2.28(5)$. Our unbiased calculation establishes that the QCP is in the universality class of the infinite-randomness fixed point predicted by previous studies based on strong disorder renormalization group technique. However, in the disordered spin-1/2 Majumdar-Ghosh chain, where a valence bond solid phase is unstable to disorder, the crossover length exponent obtained from a scaling analysis of EE disagrees with the expectation based on Imry-Ma argument. We provide a possible explanation.
\end{abstract}

\maketitle
\hypersetup{linkcolor=BrickRed}

\textit{Introduction:}
Entanglement entropy (EE) measures gross quantum mechanical correlations between different parts of a system and incorporates experimentally observable quantities in an agnostic manner.
It has recently been a subject of extensive investigation as a paradigm for understanding and classifying a wide range of quantum phases and phase transitions (QPTs)\cite{Osterloh2002,Osborne2002,Vidal2003,Gu2004,Wu2004,Refael2004,Legeza2006,Refael2007,Bounsante2007,Werlang2010,Eisert2010,Liu2016,Liu2017,Laflorencie2022,Squillante2023}. The list includes topological phenomena where it has been especially useful\cite{Kitaev2006,Levin2006,Fradkin2006,Li2008,Turner2010,Fidkowski2010,Pollman2010,Tu2013,Zaletel2013,Wen2017}, in addition to conventional symmetry breaking, impurity driven phases,\cite{Refael2004,Bonesteel2007,Refael2007} etc. Particularly appealing aspects of EE are it being a diagnostic for QPTs, a tool for extracting universal scaling behavior, and new characterizations such as the (effective) central charge\cite{Holzhey1994,Calabrese2004,Refael2004,Fradkin2006,Refael2007} and topological entanglement entropy\cite{Kitaev2006,Levin2006,Haque2007}. The extent to which it can be practically useful, in extracting quantitative properties essential to classifying the QPTs, continues to be an exciting area of research.

One dimensional spin chains with randomness have been of interest for several decades\cite{MDH1979, Bhatt1981, Bhatt1982, Bhatt1986, Fisher1994, Fisher1995, Hyman1996, Hyman1997, Yang1998, Damle2002}.
A frequent theme in these systems is the emergence of infinite randomness fixed points (IRFPs).
Pioneering works\cite{Refael2004} and subsequent studies\cite{Laflorencie2005,Chiara2006,Refael2007, Bonesteel2007,Hoyos2007,Saguia2007,Igloi2008,Refael2009,Fagotti2011,Pouranvari2013,Ruggiero2016} have established that different IRFPs can be distinguished from each other using the properties of EE, in particular via the ``effective'' central charge that controls the scaling of EE with the logarithm of size of a subsystem.\footnote{We emphasize, however, that the random critical point is not described by a conformal field theory.}
The exact results have been primarily obtained using the archetypal strong disorder renormalization group (SDRG) that becomes exact as the width of disorder distribution approaches infinity.
However, several impurity driven transitions occur at zero and finite disorder strengths where SDRG is not applicable, at least on the lattice scale. 
Hence alternate methods, that can provide independent and unbiased perspectives, are vital to exploring the properties of EE at these critical points.

In this work, we utilize the density matrix renormalization group (DMRG) that is a powerful technique for solving interacting low-dimensional systems and is recently being improved for applications in problems involving randomness\cite{Xavier2018,Torlai2018,Wybo2020,Yu2022,Wada2022,Chepiga2023} (details of our DMRG approach can be found in Appendix \ref{appendix:DMRG}).
We investigate the behavior of EE at the topological phase transition between the Haldane\cite{AKLT1987,Haldane1983b,Haldane1983c} and Random Singlet phases (RSP) in spin-1 chain.
The former is a symmetry protected topological phase that is known to be stable up to a finite disorder strength owing to its gapped nature. It transitions into the random singlet phase (RSP) upon increasing disorder beyond a critical value\cite{Hyman1996,Hyman1997,Monthus1997,Monthus1998}. The quantum critical point is proposed to be an IRFP in Refs. \cite{Hyman1997,Damle2002,Refael2007}. EE was studied in Ref. \cite{Refael2007} using SDRG technique applied to a domain wall picture and they found that it diverges with system size $L$ logarithmically, i.e. $S = \frac{c_{\rm eff}}{6} \log L + \cdots$, with $c_{\rm eff} = 1.232$. Here $L$ is the number of spins in an open chain and we take $S$ to be the mid-chain EE. Additionally, the correlation length critical exponent $\nu$ was predicted to be approximately $2.30$ in Ref. \cite{Hyman1997}.

Since this fixed point exists at an intermediate disorder strength, SDRG is not applicable for the initial steps. One may ask therefore if the critical point could be in a different universality class than the aforementioned IRFP. We investigate this question using DMRG that, as we show, can solve weak to intermediate disordered Hamiltonians if used carefully, exactly where SDRG is not trustworthy. Rather than the Heisenberg model, we study the AKLT model\cite{AKLT1987} which has a short coherence length and is exactly solvable. We add randomness such that it becomes and flows to the same RSP as the Heisenberg model at large disorder. By studying EE scaling with system size at the critical point, we obtain $c_{\rm eff} = 1.17(4)$ and correlation length exponent $\nu=2.28(5)$ that agree with the SDRG based predictions of Refs. \cite{Refael2007} and \cite{Hyman1997} respectively. This confirms that the Haldane phase to RSP transition is indeed in the same universality class as the IRFP studied in these works.

\begin{figure*}[ht!]
	\centering
	\begin{minipage}{0.6\textwidth}
		\centering
		\includegraphics[width=0.85\textwidth]{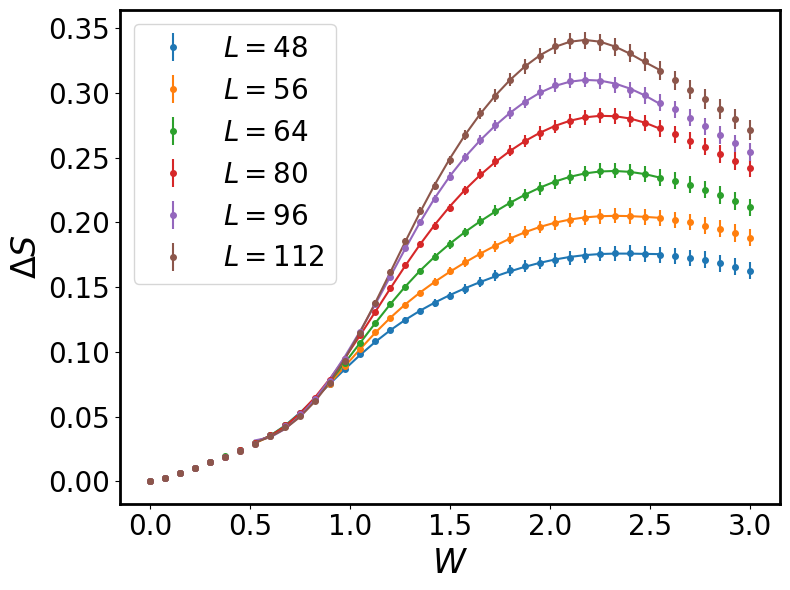}\\
		(a)
	\end{minipage}%
	\begin{minipage}{0.4\textwidth}
		\centering
		\includegraphics[width=0.8\textwidth]{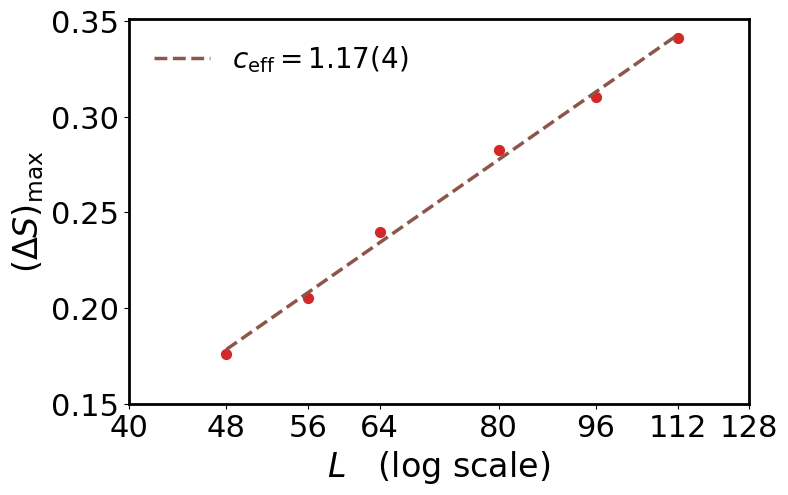}\\
		(b)\vspace{10pt}
		
		\includegraphics[width=0.8\textwidth]{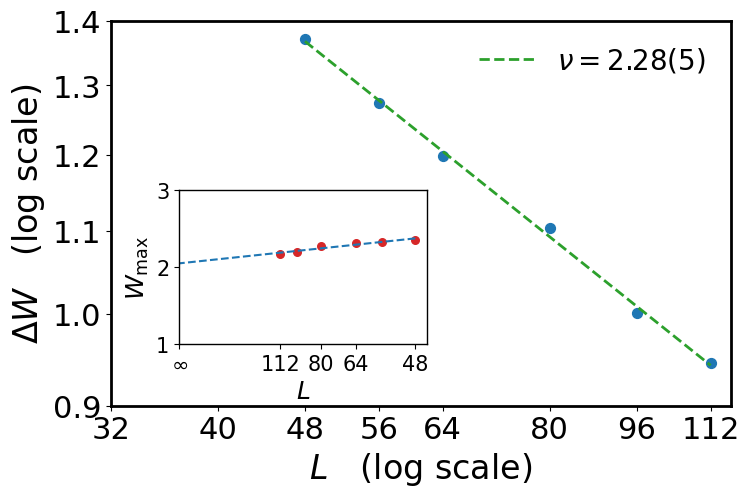}\\
		(c)
	\end{minipage}
	
	\caption{(a) Excess half-chain entanglement entropy $\Delta S = S-2\log 2$ vs. disorder strength $W$ for various chain lengths $L$ in random spin-1 chain (Eq. \eqref{eq:H_AKLT}). It is averaged over $N_D =16000$ disorder realizations except for $L=112$ where $N_D=12540$. The solid lines are fits to independent interpolating functions at each $L$. (b) A least square error fit of maxima of $\Delta S$ against $\log L$ gives effective central charge $c_{\rm eff}  = 1.17(4)$. (c) Power law scaling analysis of half-width at half-maximum on the left side of $\Delta S$, i.e. $\Delta W \sim L^{-\frac{1}{\nu}}$. Peak positions $W_{\rm max}$ are nearly independent of $L$ (inset).}
	\label{fig:AKLT}
	
\end{figure*}

As a second example of disorder induced quantum criticality, we add randomness to the Majumdar-Ghosh model\cite{MajumdarGhosh1969,Majumdar1970} such that the second-nearest neighbor anti-ferromagnetic exchange term is suppressed at large disorder and the ground state is in the same universality class as spin-1/2 RSP. Without disorder, it breaks translational invariance spontaneously by forming a dimerized state that was found to be unstable to an arbitrarily small amount of disorder in favor of an RSP phase.\cite{Yang1996} Here, we compute the mid-chain EE in its weak disorder regime and observe critical scaling confirming this expectation. Further, we obtain the crossover length scale exponent $\nu=1.16(5)$ via a scaling analysis of EE. Surprisingly, this is different than the expectation of $\nu=2$ based on an Imry-Ma\cite{ImryMa1975} type argument presented in Ref. \cite{Yang1996}.

\textit{Random biquadratic coupling AF spin-1 model:} We consider the following random biquadratic coupling model for the spin-$1$ chain:
\begin{align}
	H = \sum_{i=1}^{L-1}\left( J_i \bm S_i.\bm S_{i+1} + D_i \left(\bm S_i.\bm S_{i+1}\right)^2\right) \label{eq:H_AKLT}
\end{align}
where $\bm S_i$ correspond to the spin operators of a spin-$1$ at the $i^{\rm th}$ site. In the absence of impurities, we take $J_i = 3D_i = J$ so that one obtains the AKLT model where the ground state is in the universality class of the Haldane phase. This state can be understood as a dimerized phase of a spin-1/2 chain of length $2L$ where each spin-1 is split into two spin-$1/2$'s. The spin-$1/2$'s at the edges of an open chain are unpaired and the ground state manifold consists of four states with an energy splitting decreasing exponentially in size. Let us introduce disorder using a power law distribution of $J_i$'s:
\begin{align}
	P(J) &=  \frac{1}{W \Omega_0}\left(\frac{J}{\Omega_0}\right)^{\frac{1}{W}-1}\ \Theta\left(J(\Omega_0-J)\right),\label{eq:dis_model}\\
	D_i &= \frac{J_i^2}{3\Omega_0}\label{eq:D_J_relation}.
\end{align}
We have $\langle \log J\rangle = \log \Omega_0-W$ and the standard deviation $\sigma_{\log J} = W$, hence $W$ corresponds to the strength of disorder. Since $\Omega_0$ is an unimportant energy scale for EE calculation, we choose it by setting $\sum_i \log J_i= 0$. Importantly, Eq. \eqref{eq:D_J_relation} ensures that in $W\rightarrow 0$ limit, one recovers the AKLT model. Moreover, in $W\rightarrow \infty$ limit, $D_i \ll J_i/3$ and are irrelevant perturbations to the RSP fixed point in the spin-1 Heisenberg model as shown in appendix \ref{appendix:AKLT}.

\begin{figure*}[ht]
	\centering
	\begin{minipage}{0.5\textwidth}
		\centering
		\includegraphics[width=0.8\textwidth]{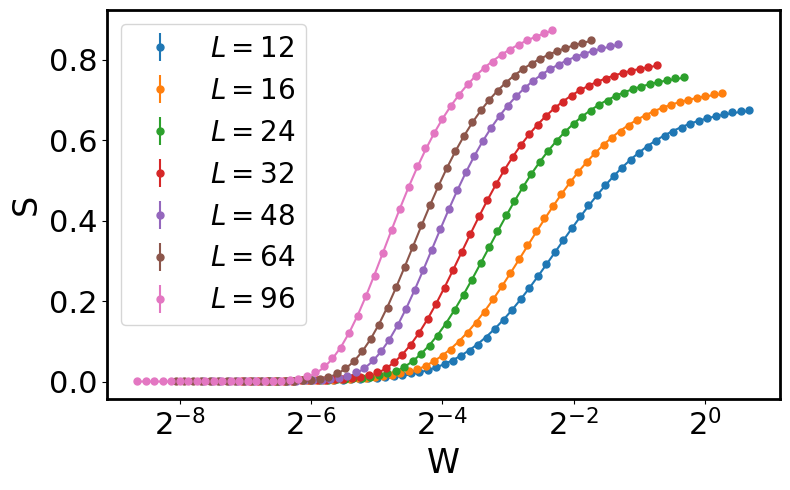}\\
		(a)
	\end{minipage}%
	\begin{minipage}{0.5\textwidth}
		\centering
		\includegraphics[width=0.8\textwidth]{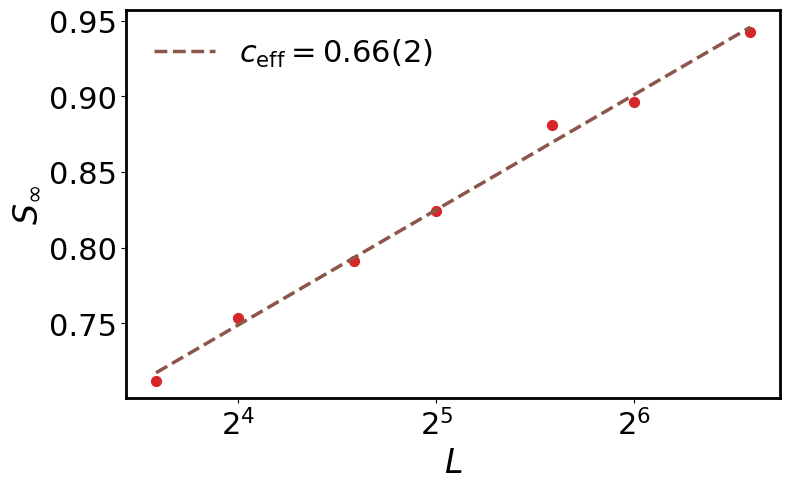}\\
		(b)
	\end{minipage}\vspace{5pt}
	\begin{minipage}{0.5\textwidth}
		\centering
		\includegraphics[width=0.8\textwidth]{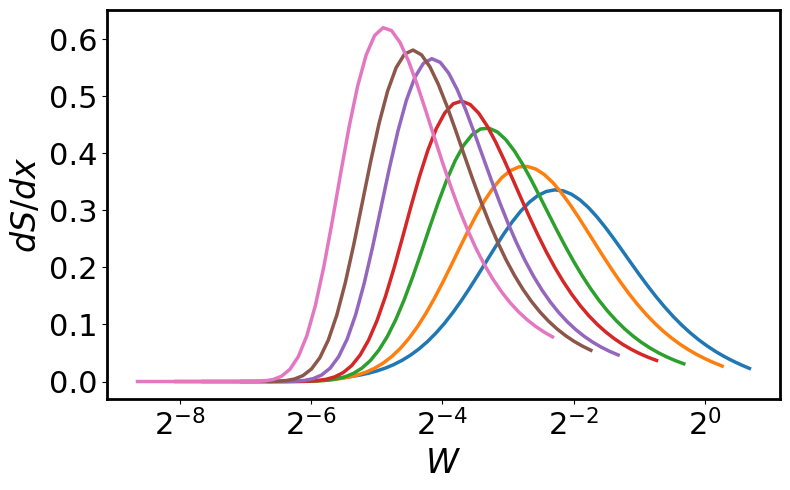}\\
		(c)
	\end{minipage}%
	\begin{minipage}{0.5\textwidth}
		\centering
		\includegraphics[width=0.8\textwidth]{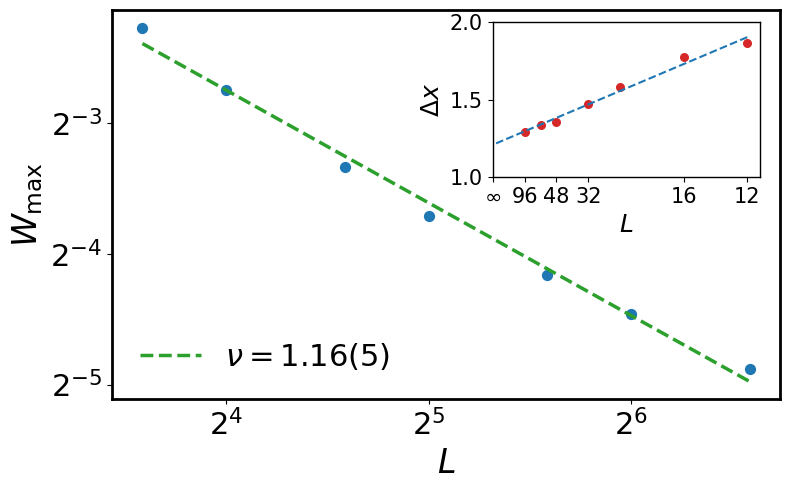}\\
		(d)
	\end{minipage}
	
	\caption{ (a) $S$ vs. $W$ for various chain lengths $L$ in random spin-1/2 model (Eq. \eqref{eq:H_MG}). (b) $S$ extrapolates to $S_\infty(L)$ in $W\rightarrow \infty$ limit from which we extract $c_{\rm eff}$. (c) $dS/dx$ obtained from the fitting functions where $x \equiv \log W$. (d) Power law scaling analysis of peak positions $W_{\rm max} = e^{x_{\rm max}}$. In the inset, full-widths at half-maximum $\Delta x$ of $dS/dx$ curves are plotted vs. $L$.}
	\label{fig:MG}
	
\end{figure*}

\textit{Critical scaling at Haldane-RSP QCP:}
In Fig. \ref{fig:AKLT}, we plot the mid-chain von-Neumann EE vs. disorder strength for various chain lengths of the model in Eq. \eqref{eq:H_AKLT}. It is defined as follows:
\begin{align}
	S = -\mathrm{Tr}\left[\rho_A\log\rho_A\right]
\end{align}
where $\rho_A$ is the density-matrix of left half chain. We assume that the ground states is an SU(2) singlet, hence EE at $W=0$ is a constant equal to $2\log 2$ that we subtract from all data-points. The excess EE, $\Delta S$ shows a clear maximum at a critical disorder strength $W=W_c$ where $W_c$ is nearly independent of size. Further, the maximum increases with $L$, and is consistent with a logarithmic divergence in $L$. Consequently, we interpret $W_c$ as the critical point between the Haldane phase and RSP. 

Our strategy for extracting $W_c$ and other critical properties is as follows. We fit the data-points at each size near the critical point to an independent interpolating function chosen to be a Gaussian envelope times a fourth order polynomial in $W$. This function is then used for further analysis as opposed to the raw data. We fit the maximum EE, ``$\Delta S_{\rm max}(L)$'', to the following expression:
\begin{align}
	\Delta S_{\rm max}(L) = \frac{c_{\rm eff}}{6} \log L + \Delta S_0. \label{eq:effective_central_charge}
\end{align}
We obtain $c_{\rm eff}=1.17(4)$. This is within two standard deviations of $c_{\rm eff} = 1.232$ predicted in Ref. \cite{Refael2007}.

We assume that the excess EE follows critical scaling in the thermodynamic limit of the form $\Delta S(W,L) \equiv  \Delta S_{\rm max}(L)f\left((W-W_c)L^{1/\nu}\right)$. This predicts that half-width at half-maximum to the left side, ``$\Delta W$'', scales as $L^{-1/\nu}$. The resulting best fit value, $\nu=2.28(5)$, is very close to the SDRG prediction, $\nu=2.3$, made in Ref. \cite{Yang1998}. We note that $\nu$ depends somewhat on the precise feature of the EE fitted. We believe this is because higher moments of the curves may not have reached the scaling limit and may require data from larger sizes. Moreover, we use only the left side of maximum. On the right side, we expect a RSP with $c_{\rm eff} = \log 3 \approx 1.099$.\cite{Refael2007} At large sizes, EE is not expected to reduce by more than $11\%$ from its critical value to the one in RSP. Moreover, DMRG simulations tend to be less reliable at large $W$ due to the wide disorder distribution at both lattice and renormalized scales.

\textit{Random second-nearest neighbor Majumdar-Ghosh spin-1/2 model:} We next consider the Majumdar-Ghosh AF spin-1/2 chain with nearest neighbor coupling J and next-nearest neighbor coupling K = J/2. We introduce disorder in both J and K, so our Hamiltonian reads-
\begin{align}
	H = \sum_{i=1}^{L-1} J_i {\bm S}_i.{\bm S}_{i+1}  + \sum_{i=1}^{L-2} K_i{\bm S}_i.{\bm S}_{i+2}.\label{eq:H_MG}
\end{align}
In the absence of disorder, $J_i = 2K_i = J$. We take $L$ to be an even number so that the ground state is unique and corresponds to singlets at bonds labeled by $2n-1/2$ where $n=1,\cdots, L/2$. The disorder distribution for $J_i$'s is same as in Eq. \eqref{eq:dis_model}. As in the previous case, $K_i$'s are chosen such that one obtains the random Heisenberg model at $W\rightarrow \infty$. In particular, we choose
\begin{align}
	K_i &= \frac{J_iJ_{i+1}}{2\Omega_0}.\label{eq:K_MG}
\end{align}
We refer the reader to Ref. \cite{Yusuf2003} and appendix \ref{appendix:AKLT} for a discussion of SDRG equations and the irrelevance of $K_i$ at the RSP in spin-1/2 Heisenberg model.

\textit{Scaling near VBS state in spin-1/2 chain:} 
There are a few subtleties in the quantitative interpretation of the scaling behavior in the spin-1/2 model. First, the RSP has power law spin-spin correlations, hence, the correlation length is infinite. We expect the transition to involve a diverging crossover length scale $\xi_{\rm cross}$ above which the VBS state disappears. Second, as we confirm below, the transition takes place at $W=0$. Therefore, any positive power of the disorder strength $W$ can serve the role of a tuning parameter making the exact value of critical exponent $\nu$ parametrization dependent. In analogy with temperature induced ferromagnet to paramagnet transition in one-dimensional pure quantum Ising model, we choose the tuning parameter to be an energy scale. The standard deviation of $J$, that has dimensions of energy, is $\approx W\Omega_0$ at $W \ll 1$. Thus we define $\nu$ via the scaling relation $W\xi_{\rm cross}^{1/\nu} = {\rm constant}$. More precisely, $\nu$ determines the scaling dimension of relevant parameter $W$ at the VBS fixed point.

In Fig. \ref{fig:MG}, we present and analyze the data for random spin-1/2 chain of Eq. \eqref{eq:H_MG}. At $W=0$, mid-chain EE is zero at the bond indexed by $(L+1)/2$. It increases monotonically with $W$ and saturates to a size dependent value. We fit our data for EE to the following functional form:
\begin{align}
	S(W,L) &= S_\infty(L)\frac{{\rm Erf}\left(\alpha(L) W + \beta(L) W^2 + \gamma(L) W^3\right)+1}{2} \label{eq:fit_func_MG}
\end{align}
where $S_\infty(L)$ is EE extrapolated to the infinite-disorder limit and $\alpha(L), \beta(L), \gamma(L)$ are fitting parameters for each size $L$. $S_\infty(L)$ is found to diverge logarithmically with $L$ as in Eq. \eqref{eq:effective_central_charge} with $c_{\rm eff} = 0.66(2)$. Ref. \cite{Refael2004} predicts $c_{\rm eff} = \log 2 \approx 0.693$ in the RSP of spin-$1/2$ chain, within two standard deviations of our numerically computed value. This agreement implies that our spin-$1/2$ model transitions from the VBS to RSP phase as disorder is introduced. Importantly, we do not observe other features in the EE suggesting that the transition is a direct one.

Based on the expectation that VBS phase is unstable to disorder, we consider the scaling ansatz $S(W,L) = f(WL^{1/\nu})$ near $W=0$ and in the thermodynamic limit. In order to extract the crossover length exponent $\nu$, we find it useful to re-parameterize the data using $x=\log W$ and transforming the scaling ansatz to following:
\begin{align}
	S(x,L) &=  F\left(x + \frac{\log L}{\nu}\right).
\end{align}
Further, the derivative of $S(x,L)$ with respect to $x$ displays a peak that is a clear feature of the transition and is reliable for a finite-size scaling analysis. Its position, i.e. $x=x_{\rm peak}$ is determined purely by the function $F(y)$ through the relation $F''(y)=0$. Hence the scaling ansatz predicts $x_{\rm peak} =  x_0-\frac{\log L}{\nu}$ using which we obtain $\nu=1.16(5)$. A consequence of our scaling assumption is that the width of $dS/dx$ curve must become independent of size in the thermodynamic limit. In the inset of Fig \ref{fig:MG}(c), we plot the full-width at half-maximum $\Delta x$ vs $1/L$. While a significant irrelevant correction is present, it nevertheless approaches a constant in the thermodynamic limit.

Our crossover length exponent is significantly different than the prediction of Ref. \cite{Yang1996} who used an argument analogous to Imry-Ma\cite{ImryMa1975} based on the domain wall picture.
Since the EE measures quantum entanglement, a quantum mechanical property, whereas the Imry-Ma argument is essentially a classical argument for the domain size, the apparent discrepancy could be argued away by positing that EE measures a quantum sub-leading length, not the domain size. We speculate this might be related to the quantum wandering of domain walls. Other possibilities include complete modification of the domain wall picture due to quantum fluctuations, or strong finite size effects for sizes studied here.


\textit{Discussion}: We have shown that ground state entanglement entropy (EE) can not only detect quantum phases and phase transitions (QPTs) in random spin chains but is also useful for characterizing their critical scaling behaviors. EE computed using DMRG in our work has provided important independent perspectives to the strong disorder renormalization group (SDRG) technique. This is especially needed because SDRG may not be valid when the quantum critical point (QCP) occurs at zero or finite disorder strength. 
Our study suggests that the DMRG technique could be used to study other one-dimensional models with disorder as a complement to other approaches such as SDRG.

At the Haldane Phase to Random Singlet Phase (RSP) transition in spin-1 chain, effective central charge and correlation length exponent extracted from the behavior of EE are consistent with previous studies.  This provides an independent and unbiased confirmation that the QCP is in the same universality class as the infinite-randomness fixed point proposed by studies based on SDRG. On the other hand, the crossover length scale exponent computed using EE in the disordered Majumdar-Ghosh model differs significantly from prior theoretical predictions\cite{Yang1996,ImryMa1975} and we discussed possible causes above. However, we leave a detailed study of this issue to future work.

\acknowledgments
We thank Kun Yang and David Huse for useful discussions.
DMRG calculations were carried out using codes written on top of TeNPy libraries\cite{tenpy}.
PK was supported by NSF through the Princeton University (PCCM) Materials Research Science and Engineering Center DMR-2011750 with additional funding received from DOE BES Grant No. DE-SC0002140 at the initial stages of the project.
RNB acknowledges support from Princeton University's UK foundation, and hospitality of the Aspen Center for Physics where this project was conceived.

\bibliography{bigbib}

\appendix
\section{DMRG ground state calculation\label{appendix:DMRG}}
In this appendix, we briefly describe our strategy for the ground state and entanglement entropy computation using DMRG. A significant obstacle to simulating ground states of random Hamiltonians is the presence of several local minima in the free energy landscape, in addition to the exponentially diverging many-body density of states such as in the RSPs. Hence a naive DMRG algorithm started from a product state fails to converge to the true ground state and usually finds a low entanglement linear combination of the low energy states. Second, roundoff errors tend to break the SU(2) symmetry of the spin-chain models considered in this work, again preferring a low-entanglement state. Our ground state search strategy is similar to the one used in Ref. \cite{Xavier2018} which provides a good initial guess to the DMRG algorithm. Essentially, a fixed set of uniformly distributed random numbers $R=\{r_i\}_{i=1}^{L-1}$ is used to create one disorder realization for each value of $W$ through the relation $J_i = r_i^W$, where $r_i \in (0,1]$. Further, we use the ground state at disorder strength $W=W_n$ as the initial guess for the ground state search at $W=W_{n+1}=W_n+\Delta W_n$ for each $R$.

The initial $W=0$ point of the two models considered in main text are exactly solvable and the ground states can be expressed as exact MPSs. Our choice of these limits was primarily motivated from the observations of Refs. \cite{Liu2016,Liu2017} who studied the disorder induced fractional quantum Hall transition problem from the Laughlin-type quantum Hall state to an insulator. In their case, the $V_1$ Haldane pseudopotential Hamiltonian, that has the Laughlin wavefunction as an exact ground state, displays relatively nice finite size scaling properties.

Our initial ground state guess is an SU(2) singlet, this implies that the net spin of the entire spin chain must stay zero over the course of DMRG algorithm. Therefore, to avoid breaking of SU(2) symmetry from roundoff errors, we add an energy penalty for non-singlet components in the matrix product state:
\begin{align}
	H_{\rm singlet} &= J_{\rm s}\left(\sum_{i=1}^L \bm S_i\right)^2,
\end{align}
where we choose $J_s=1$. Clearly $H_{\rm singlet} = J_s S_T(S_T+1)$ where $S_T$ is the net SU(2) spin of the state.

Our strategies in conjunction with the models push the maximum size and disorder strengths that can be reliably accessed in a DMRG ground state search and provide good finite size scaling properties. Nevertheless, simulating deeper into RSPs remains a challenging problem as both the ultraviolet and renormalized energy scales decrease rapidly and approach the precision of double floating point format.

\section{Irrelevance of biquadratic couplings in spin-1 Random Singlet Phase\label{appendix:AKLT}}
In this appendix, we show that the biquadratic term of the form chosen in the main text is irrelevant at the RSP fixed point in spin-1 Heisenberg model. Ref. \cite{Yang1998} derived the SDRG equations for the nearest-neighbor biquadratic spin-1  model:
\begin{align}
	H = \sum_{i=1}^{L-1} \left(J_i \bm S_i.\bm S_{i+1}+D_i \left(\bm S_i.\bm S_{i+1}\right)^2\right)
\end{align}
Upon decimating a singlet say at the bond connecting sites 2 and 3, we obtain the following effective couplings between sites 1 and 4:
\begin{align}
	\tilde J_{1,4} &= \frac{(2J_1-D_1)(2J_3-D_3)}{3 (J_2-3D_2)} - \frac{D_1 D_3}{9(J_2-D_2)}\\
	\tilde D_{1,4} &= -\frac{D_1 D_3}{9(J_2-D_2)}
\end{align}
The difference of energies between a singlet and triplet, and a singlet and quintuplet are $J-3D$, and $3J-3D$ respectively at a given bond. To make sure that only singlets are generated at SDRG, we require $3D/J<1$. If this ratio monotonically decreases under RG flow, $D$'s would be irrelevant.

Now assume that one starts with a distribution of $J_i$'s with $W\gg 1$ and $D_i = \frac{J_i^2}{3\Omega_0}$ as in the main text. We can safely assume $D_i \ll J_i/3$ at the initial point. Defining $\alpha_i = \frac{3D_i}{J_i}$, one can calculate
\begin{align}
	\tilde\alpha_{1,4} &\approx -\frac{\alpha_1\alpha_3}{18}.
\end{align}
While it fluctuates around zero, $\alpha_i$ rapidly decreases in magnitude under RG flow. Thus small biquadratic couplings of the form considered in the main text are irrelevant at the IRFP corresponding to the RSP.

\section{Effects of second-nearest neighbor exchange in spin-1/2 model\label{appendix:MG}}

In this appendix, we discuss the effects of second-nearest neighbor interactions in the random spin-$1/2$ model within a SDRG approach:
\begin{align}
	H = \sum_{i=1}^{L-1} J_i {\bm S}_i.{\bm S}_{i+1}  + \sum_{i=1}^{L-2} K_i{\bm S}_i.{\bm S}_{i+2}.
\end{align}
Let us first analyze possible effects of $K_i>0$ physically. They lead to frustration between spins separated by two-sites. Now suppose that at the current SDRG step, we decimate a singlet between sites 3 and 4. If $K_2$ and $J_4$ are sufficiently large, they would lead to a ferromagnetic coupling between spins at sites 2 and 5.
A second possibility is the formation of singlets between second-nearest neighbors.
To see this, consider three sites indexed by $i,i+1,i+2$. The three-site Hamiltonian will generally have a singlet at neighboring sites in its lowest energy state if $K_i \ll J_i, J_{i+1}$. On the other hand, if $K_i \gg J_i, J_{i+1}$, a singlet will form between second-nearest neighbors. However, under SDRG in the Heisenberg model, singlets connect only the sites that are effective nearest neighbors at some point in the RG flow and are always indexed by numbers of opposite parity. The formation of singlets between second-nearest neighbors could in principle lead to a different phase. Therefore, we demand that this must not happen for the stability of RS fixed point. Both ferromagnetic couplings and formation of second-nearest neighbor singlets can be avoided if the ratios $K_i/J_i$ and $K_i/J_{i+1}$ monotonically decrease under RG flow.

To derive SDRG equations that result from decimation of a nearest neighbor singlet, we consider 6 consecutive sites at second order perturbation theory. An antiferromagnetic coupling between neighbors 5 sites away $\tilde L_{i,i+5}$ will also be generated but will not give rise to frustration. Assume that the strongest bond connects sites 3 and 4. Using methods outlined in Ref. \cite{Hyman1996}, we can derive the following effective couplings from projecting out the corresponding singlet:
\begin{align}
	\tilde J_{1,2} &= J_1 - \frac{K_1(J_2-K_2)}{2J_3}\\
	\tilde J_{2,5} &= \frac{(J_2-K_2) (J_4-K_3)}{2J_3} \\
	\tilde J_{5,6} &= J_5 - \frac{K_4(J_4-K_3)}{2J_3}\\
	\tilde K_{1,5} &= \frac{K_1(J_4-K_3)}{2J_3}\\
	\tilde K_{2,6} &= \frac{K_4(J_2-K_2)}{2J_3}\\
	\tilde L_{1,6} &= \frac{K_1 K_4}{2J_3}
\end{align}

First, consider the RSP at infinite disorder. Let us use the model in main text $K_i = J_i J_{i+1}/2\Omega_0$. We can ignore $\tilde L_{1,6}$ since it is higher order. At the initial step and large $W$, we can assume $K_i \ll J_i$ and $K_i\ll J_{i+1}$ motivating the following approximations:
\begin{gather}
	\tilde J_{1,2} \approx J_1,\ \ \tilde J_{2,5} \approx \frac{J_2 J_4}{2J_3},\ \ \tilde J_{5,6} \approx J_5\\
	\tilde K_{1,5} \approx \frac{K_1 J_4}{2J_3},\ \ 
	\tilde K_{2,6} \approx \frac{K_4 J_2}{2J_3}
\end{gather}
We have $\frac{\tilde K_{1,5}}{\tilde J_{1,2}} \approx \frac{K_1}{J_1} \frac{J_4}{2J_3} \ll \frac{K_{1}}{J_{1}}$ and $\frac{\tilde K_{1,5}}{\tilde J_{2,5}} \approx \frac{K_1 }{J_2}$. Thus, the two ratios of $K_i$ with $J_i$ and $J_{i+1}$ either remain unchanged or reduce under RG flow. We expect that near spin-1/2 RS fixed point, $K_i$'s would be irrelevant in agreement with Ref. \cite{Yusuf2003}.

%

\end{document}